\documentclass[preprintnumbers,showpacs,floats,floatfix,amssymb,prd,twocolumn,superscriptaddress,nofootinbib]{revtex4}
\usepackage{amssymb,amsmath}
\usepackage{epsfig,hyperref}
\usepackage[svgnames]{xcolor}
\usepackage{pgf,tikz}
\usepackage{epstopdf}
\usepackage[british]{babel}
\usepackage{url}

\makeatletter
\DeclareRobustCommand*{\bfseries}{%
  \not@math@alphabet\bfseries\mathbf
  \fontseries\bfdefault\selectfont
  \boldmath
}
\makeatother

\def\be{\begin{equation}}
\def\ee{\end{equation}}
\def\beq{\begin{eqnarray}}
\def\eeq{\end{eqnarray}}

\makeatletter

\newcommand{\arXiv}[2][]{\href{http://arxiv.org/abs/#2}{\texttt{arXiv:#2\@ifempty{#1}{}{ [#1]}}}}
\makeatother

\begin{document}
\title{Mass inflation and curvature divergence near the central singularity in spherical collapse}

\author{Jun-Qi Guo}%
\email{junqi.guo@tifr.res.in}
\affiliation{Department of Astronomy and Astrophysics, Tata Institute of Fundamental Research, Homi Bhabha Road, Mumbai 400005, India}

\author{Pankaj S. Joshi}%
\email{psj@tifr.res.in}
\affiliation{Department of Astronomy and Astrophysics, Tata Institute of Fundamental Research, Homi Bhabha Road, Mumbai 400005, India}

\author{Jos\'{e} T. Galvez Ghersi}%
\email{joseg@sfu.ca}
\affiliation{Department of Physics, Simon Fraser University,\\
8888 University Drive, Burnaby, British Columbia V5A 1S6, Canada}


\begin{abstract}
  We study spherical scalar collapse toward a black hole formation and examine the asymptotic dynamics near the central singularity of the formed black hole. It is found that, in the vicinity of the singularity, due to the strong backreaction of a scalar field on the geometry, the mass function inflates and the Kretschmann scalar grows faster than in the Schwarzschild geometry. In collapse, the Misner-Sharp mass is a locally conserved quantity, not providing information on the black hole mass that is measured at asymptotically flat regions.
\end{abstract}
\maketitle

\section{Introduction}
In general relativity, due to the equivalence principle, it is challenging to define energy density associated with gravitational field. As a result, people turn to define the total mass of an isolated system. For asymptotically flat spacetime, the Arnowitt-Deser-Misner mass at spatial infinity can well describe the energy of a gravitational system. For stationary spacetime, a timelike Killing vector exists, based on which the Komar mass was defined. For nonstationary fields where there is an asymptotic time translation symmetry, one can define the Bondi energy at null infinity. Since we usually work in a nonisolated system, for convenience, several quasilocal mass definitions have been constructed~\cite{Wang_2008,Jaramillo_2008}. If we consider cases with more symmetry, it is obvious that most of these definitions become equivalent to each other and the Misner-Sharp mass is normally a well-posed one~\cite{Misner}. However, here we explore interiors of black holes in which the spacetime is dynamical and the Misner-Sharp mass function shows unusual features, testing the validity of this mass definition at such regions.

Under external perturbations, the inner horizon of a Reissner-Nordstr\"{o}m black hole can contract, and near the inner horizon the Misner-Sharp mass grows rapidly. This phenomenon is called mass inflation~\cite{Poisson_1989,Poisson_1990}. These arguments were extended to the rotating black hole case in Ref.~\cite{Barrabes_1990}. The main cause of mass inflation is that, in the vicinity of the inner horizon, due to strong gravity and repulsion, the source field changes dramatically, accumulates a large amount of energy, and then modifies the geometry significantly. To a large extent, the mass function is a geometrical quantity. As a result, the mass function diverges. In order to get more information, some numerical simulations in more realistic models have
been performed in Refs.~\cite{Gnedin_1991,Gnedin_1993,Brady_1995,Burko_1997,Burko_1997b,Hod_1997,Oren_2003,Hansen_2005}. An apparent horizon, a null, weak mass-inflation singularity along the Cauchy horizon, and a final, spacelike, central singularity were obtained. Spherical collapse of a charged scalar field with regular initial data was investigated rigorously in Ref.~\cite{Kommemi_2011}.

Besides the importance of studying dynamics near the inner horizons of charged and rotating black holes, it can be instructive to explore an even simpler case: dynamics near the central singularity of a Schwarzschild black hole. In Refs.~\cite{Burko_9711,Burko_9803}, the asymptotic dynamics near the spacelike singularity in spherical scalar collapse was studied both analytically and numerically, while mass inflation was ignored. In Ref.~\cite{Christodoulou_1991}, black hole formation was investigated analytically and a lower bound for the divergence rate of the Kretschmann scalar was obtained. In Ref.~\cite{Csizmadia}, numerical simulation of spherical scalar collapse showed that near the central singularity the mass function blows up and verified the lower bound for the divergence rate of the Kretschmann scalar reported in Ref.~\cite{Christodoulou_1991}, while analytic work on mass inflation was absent.

The purpose of this paper is to emphasize and extend our previous results on spherical scalar collapse reported in Ref.~\cite{Guo_1507}. We find that, due to the strong backreaction from the scalar field on the geometry, the mass function diverges in the vicinity of the central singularity and the Kretschmann scalar grows faster than in the Schwarzschild geometry. We argue that in collapse the Misner-Sharp mass is a locally conserved quantity, not providing information on the black hole mass that is measured at asymptotically flat regions.

This paper is organized as follows. In Sec.~\ref{sec:framework}, the framework on spherical collapse is developed. We explore mass inflation near the central singularity in Sec.~\ref{sec:mass_inflation}. The behavior of the Kretschmann scalar near the central singularity is studied in Sec.~\ref{sec:Krets_scalar}. In Sec.~\ref{sec:discussions}, conservation laws in dynamical spacetime and the behavior of the mass function for other collapses are discussed.

Throughout the paper, we set $G=c=\hbar=1$.

\section{Framework\label{sec:framework}}
Our line of thought to proceed with the calculations is not significantly different from the one we followed in Ref.~\cite{Guo_1507}. We study the effects of a scalar field on the geometry in collapse. To do so, we provide the standard definition of the energy-momentum tensor for a massless scalar field $\psi$
\be T^{(\psi)}_{\mu\nu}
=\psi_{,\mu}\psi_{,\nu}-\frac{1}{2}g_{\mu\nu}g^{\alpha\beta}\psi_{,\alpha}\psi_{,\beta}.\label{energy_tensor_psi}\ee
In addition to this, we use the double-null coordinates,
\be
\begin{split}
ds^{2} &= -4e^{-2\sigma}dudv+r^2(d\theta^2+\sin^2{\theta}d\phi^2)\\
&= e^{-2\sigma}(-dt^2+dx^2)+r^2(d\theta^2+\sin^2{\theta}d\phi^2),
\end{split}
\label{double_null_metric}
\ee
where $\sigma$ and $r$ depend on coordinates $t$ and $x$, $u=(t-x)/2$, and $v=(t+x)/2$. Thus, we can find the dynamical equations just as in Ref.~\cite{Guo_1507}:
\be r(-r_{,tt}+r_{,xx})-r_{,t}^2+r_{,x}^2 = e^{-2\sigma},\label{equation_r}\ee
\be -\sigma_{,tt}+\sigma_{,xx} + \frac{r_{,tt}-r_{,xx}}{r}+4\pi(\psi_{,t}^2-\psi_{,x}^2)=0,\label{equation_sigma}\ee
\be -\psi_{,tt}+\psi_{,xx}+\frac{2}{r}(-r_{,t}\psi_{,t}+r_{,x}\psi_{,x})=0. \label{equation_psi}\ee
Constraints emerge from $\{uu\}$ and $\{vv\}$ components of the Einstein equations,
\be r_{,uu}+2r_{,u}\sigma_{,u}+4\pi r\psi_{,u}^2=0, \label{constraint_eq_uu}\ee
\be r_{,vv}+2r_{,v}\sigma_{,v}+4\pi r\psi_{,v}^2=0. \label{constraint_eq_vv}\ee
For numerical stability concerns, we use the constraint equation (\ref{constraint_eq_uu}) instead of Eq.~(\ref{equation_sigma})~\cite{Frolov_2004}. Defining a new variable $g$ as $g\equiv-2\sigma-\ln(-r_{,u})$, we can recast Eq.~(\ref{constraint_eq_uu}) as the equation of motion for $g$,
\be g_{,u}=4{\pi}\cdot\frac{r}{r_{,u}}\cdot\psi_{,u}^2. \label{equation_g}\ee

We impose $r_{,tt}=r_{,t}=\sigma_{,t}=\psi_{,t}=0$ and $\psi(r)=a\cdot\tanh\left[(r-r_0)^2\right]$ at $t=0$ to be our initial conditions with $a=0.1$ and $r_{0}=5$. The Misner-Sharp mass $m$ can be read from its original definition in Ref.~\cite{Misner},
\be g^{\mu\nu}r_{,\mu}r_{,\nu}=e^{2\sigma}(-r_{,t}^2+r_{,x}^2){\equiv}1-\frac{2m}{r}.\label{mass_definition}\ee
We require $r=m=g=0$ at the origin $(x=0,t=0)$. The parameters $r$, $m$, and $g$ on the initial slice $(x=x,t=0)$ are determined by numerically integrating the equations below along the line bounded by $x=0$ and $x=x_{b}$ (see~\cite{Frolov_2004,Guo_1312} for more details):
\begin{align}
r_{,x}&=\left(1-\frac{2m}{r}\right)e^g,\nonumber\\
\nonumber\\
m_{,r}&=4{\pi}r^2\left[V+\frac{1}{2}\left(1-\frac{2m}{r}\right)\psi_{,r}^2\right],\nonumber\\
\nonumber\\
g_{,r}&=4{\pi}r\psi_{,r}^2.\nonumber
\end{align}
From Eqs.~(\ref{equation_r})-(\ref{equation_psi}), it is enough to use a second-order Taylor expansion to determine the values of $r$, $\sigma$, and $\psi$ at $t={\Delta}t$. The value of $g$ at $t={\Delta}t$ is determined by the definition of $g$. The spatial range is $x\in[0;\:10]$. Set $r=\psi_{,x}=0$ at $x=0$. The value of $g$ at $x=0$ and the outer boundary conditions are obtained via extrapolation. The spatial and temporal grid spacings are ${\Delta}x={\Delta}t=0.005$. The finite difference method and leapfrog integration scheme are implemented. The numerical code is second-order convergent. Further details are described in Ref.~\cite{Guo_1507}.

\begin{figure*}[t!]
  \epsfig{file=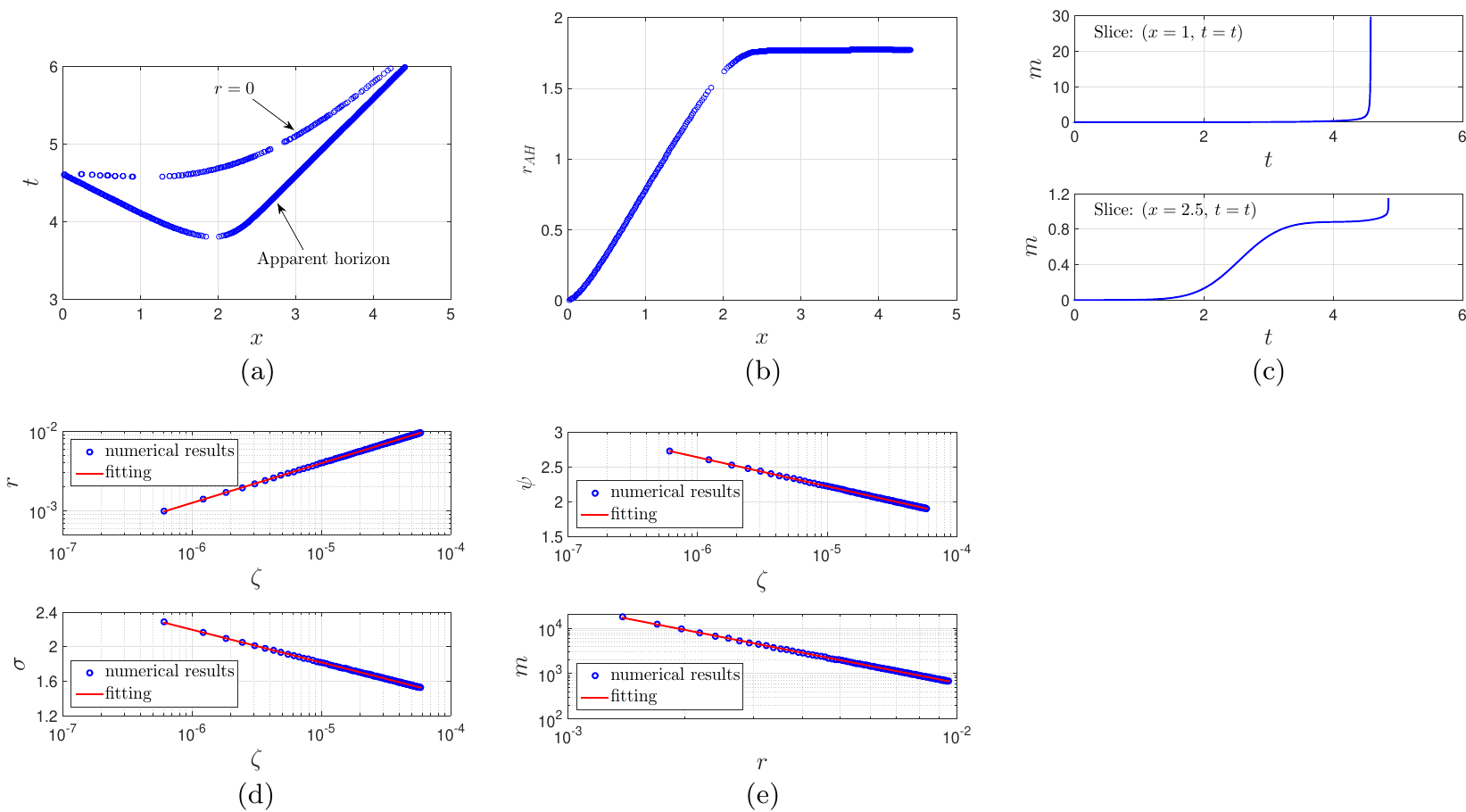, width=0.94\textwidth}
  \caption{Solutions near to the central singularity in neutral scalar collapse.
  (a) and (b): apparent horizon and singularity curve of the formed black hole.
  (c) Misner-Sharp mass function along the slices $(x=1,t=t)$ and $(x=2.5,t=t)$.
  (d) and (e) are results for the slice $(x=1,t=t)$.
  (d) $\ln r=a\ln\zeta+b$, $a=0.49948\pm0.00004$, $b=0.2172\pm0.0005$.
      $\sigma=a\ln\zeta+b$, $a=-0.1651\pm0.0001$, $b=-0.083\pm0.001$.
  (e) $\psi=a\ln\zeta+b$, $a=-0.18129\pm0.00003$, $b=0.1322\pm0.0003$.
      $\ln m=a\ln r+b$, $a=-1.670\pm0.002$, $b=-1.258\pm0.009$.}
  \label{fig:solutions_collapse_x_1}
\end{figure*}

\section{Mass inflation near the central singularity\label{sec:mass_inflation}}
The code for collapse is run and a black hole is formed. The apparent horizon of the formed black hole has a radius of $r_{\scriptsize{\mbox{AH}}}\approx1.8$. We plot the apparent horizon and singularity curve $r(x,t)=0$ in Figs.~\ref{fig:solutions_collapse_x_1}(a) and \ref{fig:solutions_collapse_x_1}(b). The Misner-Sharp mass function along the slices $(x=1,t=t)$ and $(x=2.5,t=t)$ is shown in Fig.~\ref{fig:solutions_collapse_x_1}(c), from which one can see that the mass function diverges near the central singularity.

\begin{figure}[t!]
  \epsfig{file=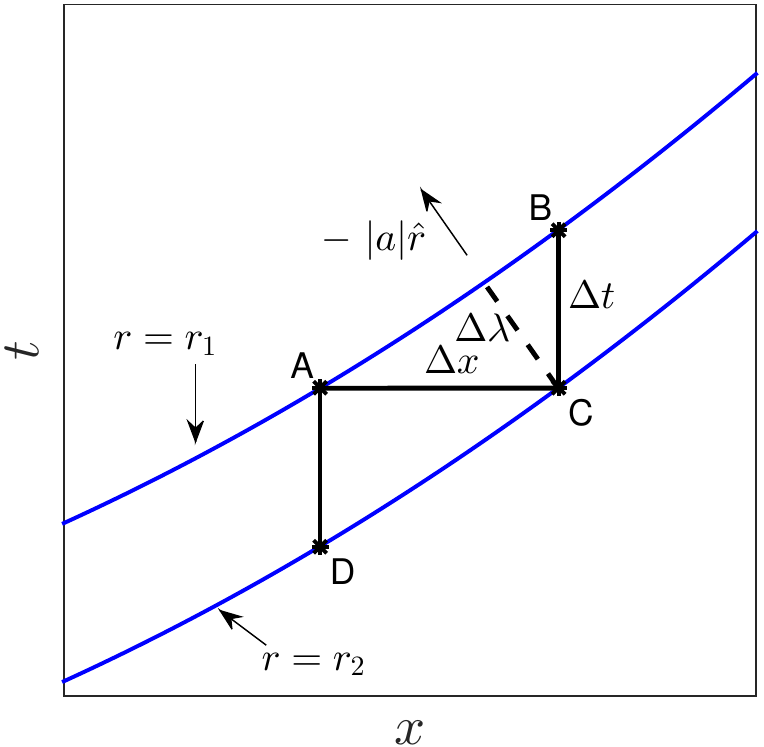, width=5cm}
  \caption{Spatial vs temporal variations near the central singularity in spherical collapse. $\Delta\lambda$ and $-|a|\hat{r}$ are orthogonal to the $r=r_{1}$ and $r=r_{2}$ constant curves. The value of $a$ is such that the magnitude of $-|a|\hat{r}$ has the length as shown in the figure.}
  \label{fig:spatial_vs_temporal}
\end{figure}

As discussed in Ref.~\cite{Guo_1312}, in the vicinity of the central singularity, the ratios between the spatial and temporal derivatives of the quantities $r$, $\sigma$, and $\psi$ take similar values and are determined by the slope of the singularity curve. With this feature and our numerical results, the Einstein equations and the equation of motion for $\psi$~(\ref{equation_r})-(\ref{equation_psi}) can be reduced as follows:
\begin{align}
rr_{,tt}&\approx-r_{,t}^2,\label{equation_r_asymptotic}\\
\nonumber\\
\sigma_{,tt}&\approx\frac{r_{,tt}}{r}+4\pi\psi_{,t}^2,\label{equation_sigma_asymptotic}\\
\nonumber\\
\psi_{,tt}&\approx-\frac{2}{r}r_{,t}\psi_{,t}.\label{equation_psi_asymptotic}
\end{align}
The asymptotic solutions to Eqs.~(\ref{equation_r_asymptotic})-(\ref{equation_psi_asymptotic}) are~\cite{Guo_1507,Guo_1312}
\begin{align}
r&\approx A\zeta^{\beta},\label{r_asymptotic}\\
\nonumber\\
\sigma&\approx B\ln\zeta+\sigma_0,\label{sigma_asymptotic}\\
\nonumber\\
\psi&\approx C\ln\zeta,\label{psi_asymptotic}
\end{align}
where $\beta\approx1/2$ and $B\approx\beta(1-\beta)-4{\pi}C^2$. $\zeta=t_0-t$, and $t_0$ stands for the coordinate time on the singularity curve. In Fig.~\ref{fig:spatial_vs_temporal}, $\zeta$ means the segments of $AD$ and $BC$. Numerical and fitting results for $r$ and $\sigma$, and $\psi$ on the slice $(x=1,t=t)$ are plotted in Figs.~\ref{fig:solutions_collapse_x_1}(d) and \ref{fig:solutions_collapse_x_1}(e), respectively.

Combining Eqs.~(\ref{mass_definition}), (\ref{r_asymptotic}), and (\ref{sigma_asymptotic}), we are able to rewrite the mass function as
\be
\begin{split}
m&=\frac{r}{2}[1+e^{2\sigma}(r_{,t}^2-r_{,x}^2)]\\
&\approx\left[\frac{1}{8}(1-J^2)A^{3}e^{2\sigma_0}\right]\zeta^{-D^2}\\
&\approx\left[\frac{1}{8}(1-J^2)A^{3+2D^2}e^{2\sigma_0}\right]r^{-2D^2},
\end{split}
\label{mass_analytic}
\ee
where $J(\equiv{r_{,x}/r_{,t}})$ is the steepness of the singularity curve and $D\equiv\sqrt{8\pi}C$. As shown in Eq.~(\ref{mass_analytic}), the mass function inflates near the central singularity. Our results from numerical integration and fitting for $m$ are plotted in Fig.~\ref{fig:solutions_collapse_x_1}(e). Note that in Ref.~\cite{Csizmadia} the divergence of the mass parameter $m$ near the central singularity was also reported based on numerical simulation, while analytic work was absent.

Furthermore, we examine the differential form of the mass function. Using Eq.~(\ref{mass_definition}) and Einstein equations, one obtains~\cite{Poisson_1989,Poisson_1990}
\be \frac{\partial m}{\partial x^a}=4{\pi}r^{2}[T^{b}_{a}-\delta^{b}_{a}T^{(2)}]\frac{\partial r}{\partial x^b},\label{dm_dx}\ee
where $x^a$ represents the coordinates $t$ and $x$. The trace of the energy-momentum tensor on these indices $T^{(2)}({\equiv}T^{t}_{t}+T^{x}_{x})$ cancels for the case of a massless scalar field $\psi$. Unit vectors along the radial, temporal, and spatial directions are denoted by $\hat{r}$, $\hat{t}$, and $\hat{x}$, respectively. $\hat{r}$ is normal to the contour lines $r=\mbox{const}$, as shown in Fig.~\ref{fig:spatial_vs_temporal}. Combination of Eqs.~(\ref{energy_tensor_psi}) and (\ref{dm_dx}) yields the gradient of $m$
\be
\begin{split}
\nabla{m}
&=\frac{\partial m}{\partial t}\hat{t}+\frac{\partial m}{\partial x}\hat{x}\\
&=-4{\pi}r^{2}\cdot\frac{1}{2}e^{2\sigma}(\psi_{,t}^2+\psi_{,x}^2)[r_{,t}\hat{t}+r_{,x}(-\hat{x})]\\
&=-4{\pi}r^{2}\cdot\frac{1}{2}e^{2\sigma}(\psi_{,t}^2+\psi_{,x}^2)r_{,t}\sqrt{1+J^2}(-\hat{r})\\
&\hphantom{i}{\approx} \frac{{\Delta}m}{{\Delta}\lambda}(-\hat{r}).
\end{split}
\label{dm_dlambda}\ee
From Fig.~\ref{fig:spatial_vs_temporal}, one obtains
\be \Delta\lambda
=\frac{\Delta t\cdot\Delta x}{\sqrt{(\Delta t)^2+(\Delta x)^2}}
\approx\frac{\Delta t}{\sqrt{1+J^2}}.\nonumber\ee
Note that
\be \frac{\Delta r}{\Delta\lambda}
=\frac{\Delta r}{\Delta t}\cdot\frac{\Delta t}{\Delta \lambda}
{\approx}r_{,t}\sqrt{1+J^2},\nonumber\ee
then we have
\be m_{,r}\approx\frac{{\Delta}m}{{\Delta}r}
=\frac{{\Delta}m}{{\Delta}\lambda}\cdot\frac{{\Delta}\lambda}{{\Delta}r}
\approx-4{\pi}r^{2}\cdot\frac{1}{2}e^{2\sigma}(\psi_{,t}^2+\psi_{,x}^2).
\label{dm_dr}\ee
Equation~(\ref{dm_dr}) gives the effective energy density for a scalar field in the neighborhood of the central singularity,
\be
\rho_{\scriptsize{\mbox{ef{}f}}}
=\frac{1}{2}e^{2\sigma}(\psi_{,t}^2+\psi_{,x}^2)
{\sim}r^{-3-2D^2}.
\ee
This is consistent with the Misner-Sharp mass definition~(\ref{mass_analytic}).

There are noticeable similarities between the central singularity in a Schwarzschild black hole and the inner horizon in a Reissner-Nordstr\"{o}m black hole during mass inflation. The Misner-Sharp mass function is expressed by the metric components. For static Schwarzschild and Reissner-Nordstr\"{o}m black holes, the Misner-Sharp mass function remains constant throughout the whole spacetime. For the cases of collapse studied in Schwarzschild and Reissner-Nordstr\"{o}m geometries, when the backreaction of the scalar field is neglected, the geometry does not change and the Misner-Sharp mass function stays fixed. When the backreaction is taken into account, the geometry is effectively modified by the scalar field and mass inflation occurs.

\begin{figure*}[t!]
  \epsfig{file=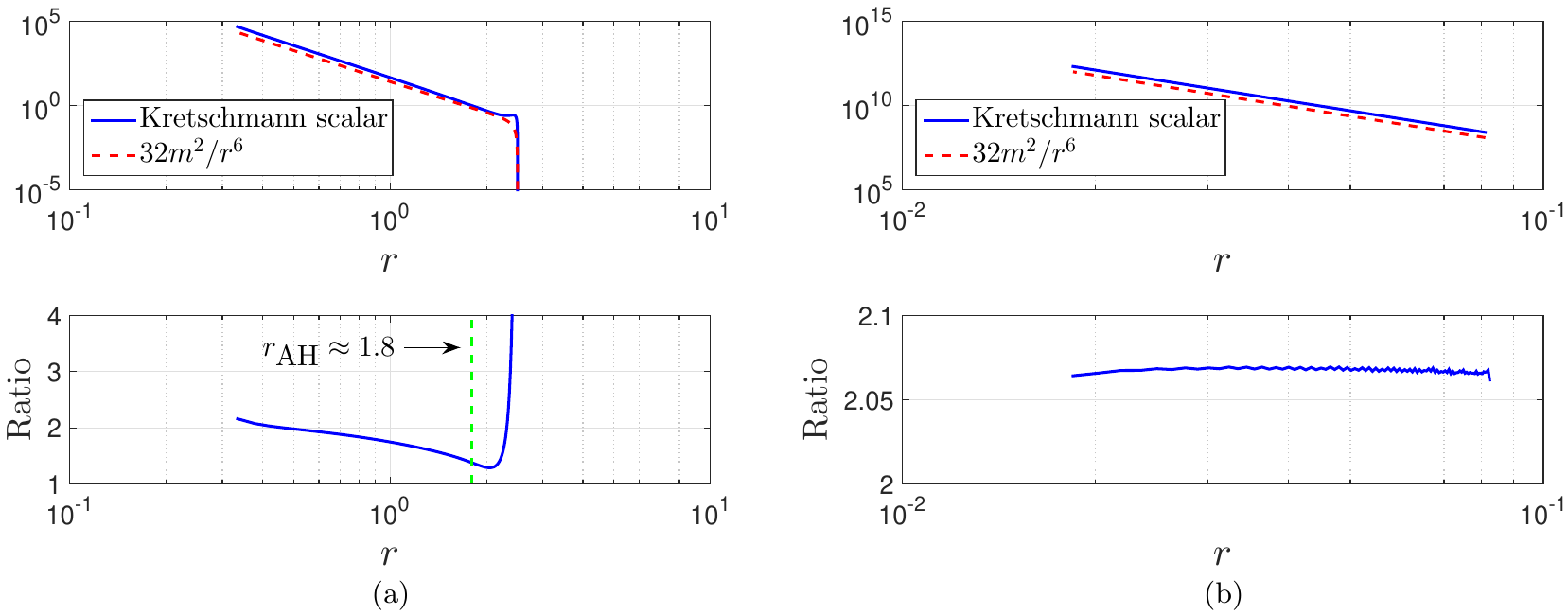, width=0.8\textwidth}
  \caption{Verification of Eq.~(\ref{lower_bound_K}): $K{\geq}32m^2/r^6$ for $r<r_{\scriptsize{\mbox{AH}}}$. The sample slice is $(x=2.5,t=t)$. (a) results for the time range $t\in[0;\:t_{0}]$ with grid spacings ${\Delta}x={\Delta}t=0.005$, where $t_{0}$ is the coordinate time for $r=0$. (b) results near $r=0$ obtained via mesh refinement. In the figure, $\mbox{Ratio}{\equiv}K/(32m^2/r^6)$.}
  \label{fig:K_scalar}
\end{figure*}

\section{Divergence of the Kretschmann scalar near the central singularity\label{sec:Krets_scalar}}
In Ref.~\cite{Christodoulou_1991}, spherical scalar collapse was explored analytically, and a lower bound for the divergence rate of the Kretschmann scalar $K[{\equiv}R_{\mu\nu\alpha\beta}R^{\mu\nu\alpha\beta}]$ inside the apparent horizon was obtained,
\be K\geq\frac{32m^2}{r^6} \hphantom{ddd} \mbox{for } r<r_{\scriptsize{\mbox{AH}}}.\label{lower_bound_K}\ee
A numerical verification of this expression was provided in Ref.~\cite{Csizmadia}. Here, besides verifying this result, we derive approximate analytic expression for the Kretschmann scalar near the central singularity.

In the double-null coordinates~(\ref{double_null_metric}), the Kretschmann scalar can be written as
\be
\begin{split}
K=
&\frac{4e^{4\sigma}}{r^4}[e^{-4\sigma}+r^{4}(\sigma_{,tt}-\sigma_{,xx})^2\\
&+2e^{-2\sigma}(r_{,t}^2-r_{,x}^2)+(r_{,t}^2-r_{,x}^2)^2\\
&+2r^2(r_{,xx}^2+r_{,tt}^2-2r_{,xt}^2)\\
&+4r^2(r_{,t}^2\sigma_{,t}^2+r_{,x}^2\sigma_{,x}^2-r_{,t}^2\sigma_{,x}^2-r_{,x}^2\sigma_{,t}^2)\\
&+4r^{2}(r_{,xx}r_{,x}\sigma_{,x}+r_{,tt}r_{,t}\sigma_{,t}+r_{,tt}r_{,x}\sigma_{,x}\\
&+r_{,xx}r_{,t}\sigma_{,t}-2r_{,xt}r_{,t}\sigma_{,x}-2r_{,xt}r_{,x}\sigma_{,t})].
\end{split}
\label{K_scalar}\ee
We verify the relation~(\ref{lower_bound_K}) on several slices $(x=x_{i},t=t)$ with $0.5{\leq}x_{i}{\leq}4$. The results for a sample slice $(x=2.5,t=t)$ are plotted in Fig.~\ref{fig:K_scalar}.

In spite of the large expression in~(\ref{K_scalar}), this can be simplified in the vicinity of the central singularity, where the metric is expressed by the Kasner solution~\cite{Guo_1312},
\be
ds^2=-d\tau^2+\sum\limits_{i=1}^3 \tau^{2p_i}dx_{i}^{2}.
\label{Kasner_solution}
\ee
The proper time $\tau$ is obtained via
\be \tau=\int_0^{\zeta}e^{-\sigma}d\zeta\approx\frac{4}{3+2D^2}\zeta^{\frac{3+2D^2}{4}}.\label{proper_time}\ee
$dx_{1}$, $dx_{2}$, and $dx_{3}$ correspond to $dx$, $d\theta$, and $\sin{\theta}d\phi$ in the double-null coordinates~(\ref{double_null_metric}), respectively. $p_1=(-1+2D^2)/(3+2D^2)$, and $p_2=p_3=2/(3+2D^2)$. $q[=4D/(3+2D^2)]$ represents the scalar field contribution, $\psi=q\ln\tau$. The Kasner exponents satisfy $p_1+p_2+p_3=1$ and $p^{2}_{1}+p^{2}_{2}+p^{2}_{3}=1-q^2$. In the Kasner metric~(\ref{Kasner_solution}), using Eqs.~(\ref{r_asymptotic}) and (\ref{proper_time}), the Ricci and Kretschmann scalars can be expressed as
\be
\begin{split}
R&=\frac{6D^2(-3+D^2)}{(3+2D^2)^2}\tau^{-2}\\
&\approx\frac{3D^2(-3+D^2)}{8}\left(\frac{r}{A}\right)^{-3-2D^2},
\end{split}
\label{Ricci_scalar}\ee
\be
\begin{split}
K&=\frac{12(16-40D^2+99D^4-26D^6+3D^8)}{(3+2D^2)^4}\tau^{-4}\\
&\approx\frac{3(16-40D^2+99D^4-26D^6+3D^8)}{64}\left(\frac{r}{A}\right)^{-2(3+2D^2)}.
\end{split}
\label{K_scalar_Kasner}\ee
The expressions~(\ref{Ricci_scalar}) and (\ref{K_scalar_Kasner}) are consistent with those obtained in Schwarzschild-like coordinates under quasihomogeneity assumption in Ref.~\cite{Burko_9803}. $H(\equiv16-40D^2+99D^4-26D^6+3D^8)$ is positive for arbitrary values of $D$ and has a minimum value of $11.7$ at $D\approx0.47$. When $D=0$, the Schwarzschild metric is recovered~\cite{Guo_1312}, and Eqs.~(\ref{Ricci_scalar}) and (\ref{K_scalar_Kasner}) respectively lead to $R=0$ and $K=48m^2/r^6$, as expected. With Eq.~(\ref{mass_analytic}), there is
\be \frac{32m^2}{r^6}\approx\frac{(1-J^2)^{2}e^{4\sigma_0}}{2}\left(\frac{r}{A}\right)^{-2(3+2D^2)}.\label{lower_bound_m}\ee
Then $K$ and $32m^2/r^6$ are at the same scale. Equations~(\ref{lower_bound_K}), (\ref{K_scalar_Kasner}), and (\ref{lower_bound_m}) show that the backreaction of the scalar field enhances the growth of the Kretschmann curvature compared to the Schwarzschild case by a factor of $r^{-4D^2}$.

\section{Discussions\label{sec:discussions}}
In collapse, in the vicinities of the central singularity of a Schwarzschild black hole and the inner horizons of Reissner-Nordstr\"{o}m and Kerr black holes, the dynamics is \emph{local}. The quantity $m$ is just a parameter which varies at each point, not giving \emph{global} information on the black hole mass. One may interpret this issue via an analog to Newtonian gravity. In Newtonian gravity, suppose we want to measure the mass $M$ of a source sphere with a pointlike test mass $m$ using a torsion balance. Denote the distance between the two masses by $r$. Without perturbations, the gravitational force between the two masses is simply $F=GMm/r^2$. However, if another mass $\delta M$ passes by the test mass, the gravitational field near the test mass can become very strong and local and is not able to provide accurate information on the source sphere. When the perturbation mass is gone or sticks to the source sphere, one can measure the mass of the source object accurately again.

In the case of static stars, the mass function is a globally defined charge emerging from Killing vectors, which work as generating currents.
Now we argue that the locally defined Misner-Sharp mass cannot be promoted to a global quantity during collapse. ``Kodama's miracle'' that was revealed for the first time in Ref.~\cite{Kodama:1979vn} provides a set of conservation laws required for the study of local dynamics in a spherically symmetric spacetime. Conserved charges might be found from Kodama's vector flow, which is described by $K^a=-\epsilon^{ab}r_{,b}$, where $(a,b)\rightarrow(T,r)$ and $\epsilon^{ab}$ is the two-dimensional Levi-Civita tensor for a curved spacetime. The Misner-Sharp mass as defined in Eq.~(\ref{mass_definition}) is an example of a conserved charge from the current $J^a=G^{ab}K_b$. The last conserved current is equivalent to Eq.~(\ref{dm_dx}) for the case of a nonvacuum solution in general relativity. We provide a simplified argument to show that it is only in the static case when nonlocal currents can be defined. These currents are the so-called Killing vectors, which are given by $\nabla_\nu\xi_\mu+\nabla_\mu\xi_\nu=0$. It is always possible to reparametrize the null generators in a way that allows us to write $\xi^\mu=(1,\:0,\:0,\:0)$, oriented with the ``temporal'' direction. Note that $\nabla_\mu\xi_\nu+\nabla_\nu\xi_\mu=g_{\mu\nu,0}$. Therefore, if the reparametrized metric is independent of the ``zeroth'' coordinate, the Killing condition $(\nabla_\nu\xi_\mu+\nabla_\mu\xi_\nu=0)$ is fulfilled and $\xi^\mu$ can be called a Killing vector. Hence, it is not possible to align the null generators with $\xi^\mu$ when the spacetime becomes dynamical.

To study the collapse case, we consider a minor modification of the usual Schwarzschild coordinates, coherent with the discussions in Refs.~\cite{Bengtsson,Kodama:1979vn,Abreu:2010ru},
\beq ds^2&=&-e^{2\lambda}\left(1-\frac{2m}{r}\right)dT^2+\frac{dr^2}{1-\frac{2m}{r}}+r^2d\Omega^2\nonumber\\
&=&\left(1-\frac{2m}{r}\right)d\tau^2+\frac{dr^2}{1-\frac{2m}{r}}+r^2d\Omega^2,
\nonumber
\eeq
where $m=m(T,r)$, $\lambda=\lambda(T,r)$, and the last redefinition is possible only if $\lambda$ depends on $T$ exclusively, in a way that $d\tau/d T = e^{\lambda}$. Henceforth, the Kodama vector $K^a[=-\epsilon^{ab}r_{,b}=-(1,\:0,\:0,\:0)]$ is aligned with our previous prescription of $\xi^\mu$. Nonetheless, $m$ being a time-dependent parameter, this is an example in which Kodama currents do not satisfy the Killing condition. Therefore, all the local
charges, $Q=\int J_ad\Sigma^a$, obtained by integration on a fixed hypersurface $\Sigma$ can be defined globally only in the stationary case. Consequently, the aforementioned divergence of the mass function must not be considered as a nonlocal phenomenon, which would be more relevant. However, it is worthwhile to mention that time dependence is not an obstacle in order to find conserved quantities in the neighborhood of a point where the currents are well defined. It is widely known that the greatest achievement of the local approach is to hold the invariance of these quantities even in dynamical situations. (See Refs.~\cite{Kodama:1979vn,Abreu:2010ru} for explicit details.) Furthermore, it is not required to have complete knowledge of the whole evolution history to provide these quantities. Such a peculiarity can also represent a limitation under local (time and position) displacements of the mass function, affecting the goal of a unique mass parameter. In fact, as shown in Eq.~(\ref{mass_analytic}), $m$ is a function of $r$ and $D$. As demonstrated in Ref.~\cite{Guo_1312}, the parameter $D$, which represents the strength of the scalar field, varies slowly along the singularity curve.

Mass inflation emerges from the backreaction of the scalar field on the geometry. A massless scalar field is also called \lq\lq stiff matter\rq\rq~with equation of state $p=\rho$, where $p$ and $\rho$ denote the pressure and energy density, respectively. In fact, not every type of matter field can have such a strong backreaction so as to modify the geometry. In perfect fluid collapse with equation of state $p=k\rho$ where $0\:{\leq}\:k<1$, the dynamics near a spacelike singularity remains the same as in vacuum up to the second order~\cite{Belinsky:1988mc,Belinski_1404}. Considering spherical symmetry, spacetime geometry is given by the Schwarzschild solution and the Misner-Sharp mass becomes a constant. The Lema\^{i}tre-Tolman-Bondi dust collapse model is a typical example, in which the mass function is well defined and finite all the way up to the spacetime singularity~\cite{Joshi:1993zg}.

Now we summarize our results. First, besides the inner horizons of charged and rotating black holes, it was found that mass inflation also takes place in the vicinity of the central singularity of a neutral black hole. Secondly, it was argued that, for a collapsing system, the quasilocal Misner-Sharp mass definition cannot be extended to describe the global dynamics inside the black hole. Thirdly, we obtained approximate analytic expression for the Kretschmann scalar near the central singularity and found that the scalar field backreaction makes the Kretschmann scalar grow faster than in the Schwarzschild geometry.
\\

\section*{ACKNOWLEDGMENTS}
The authors thank Lior M. Burko, Andrei V. Frolov, Ist\'{v}an R\'{a}cz, and Alex Zucca for the useful discussions. The authors are grateful to the referees for the helpful comments. This project has been partly funded by the Discovery Grants program of the Natural Sciences and Engineering Research Council of Canada.


\end{document}